\documentclass[12pt]{article}
\usepackage{lineno}
\usepackage{graphicx}
\usepackage{url}
\usepackage{color}
\usepackage{multirow}
\usepackage[hmargin=1in,vmargin=1in]{geometry}

\pdfoutput=1

\usepackage{upgreek}
\usepackage{eurosym}
\usepackage{fancyhdr}
\usepackage{units}
\usepackage{xspace}
\usepackage{siunitx}

\newcommand{\nova}{\mbox{NO$\upnu$A}\xspace}

\newlength{\captwidth}

\date{}

\begin{document}

\begin{centering}

\bigskip\bigskip\bigskip\bigskip
\vspace{10cm}

{\LARGE {\bf RADAR - R\&D Argon Detector at Ash River \\ \vskip 0.2 cm
- Letter of Intent -} }

\bigskip\bigskip\bigskip\bigskip

P.~Adamson$^{1}$, S.~Agarwalla$^{2}$, A.~Aurisano$^{3}$, J.J.~Evans$^{4}$, P.~Guzowski$^{4}$, A.~Habig$^{5}$, A.~Holin$^{6}$, J.~Huang$^{7}$, M.~Kordosky$^{8}$, A.~E.~Kreymer$^{1}$, K.~Lang$^{7}$, M.~Marshak$^{9}$, R.~Mehdiyev$^{7}$, W.~H.~Miller$^{9}$, D.~Naples$^{10}$, R.J.~Nichol$^{6}$, V.~Paolone$^{10}$, R.~B.~Patterson$^{11}$, A.~Sousa$^{3}$, J.~Thomas$^{6}$, L.~H.~Whitehead$^{6}$\\

\bigskip\bigskip

{\small \em
$^{1}$Fermi National Accelerator Laboratory, Batavia, Illinois~60510, USA\\
$^{2}$Institute of Physics, Sachivalaya Marg, Bhubaneswar, Odisha~751005, India.\\
$^{3}$Department of Physics, University of Cincinnati, Cincinnati, Ohio~45221, USA.\\
$^{4}$School of Physics and Astronomy, University of Manchester, Oxford Road, Manchester M13~9PL, United Kingdom.\\
$^{5}$Physics Department, University of Minnesota Duluth, 10 University Dr., Duluth, Minnesota~55812, USA\\
$^{6}$Department of Physics and Astronomy, University College London, Gower Street, London~WC1E~6BT, United Kingdom.\\
$^{7}$Department of Physics, University of Texas at Austin, 1 University Station C 1600, Austin, TX~78712, USA\\
$^{8}$Department of Physics, College of William \& Mary, Williamsburg, Virginia~23187, USA.\\
$^{9}$University of Minnesota, Minneapolis, Minnesota~55455, USA\\
$^{10}$University of Pittsburgh, Pittsburgh, Pennsylvania~15213, USA\\
$^{11}$Division of Physics, Mathematics, and Astronomy, California Institute of Technology, Pasadena, California~91125, USA.\\
}

\bigskip\bigskip

\today

\end{centering}

\abstract{In the RADAR project described in this Letter of Intent, we propose to deploy a 6\,kton liquid argon TPC at the \nova Far detector building in Ash River, Minnesota, and expose it to the NuMI beam during \nova running. It will significantly add to the physics capabilities of the \nova program while providing LBNE with an R\&D program based on full-scale TPC module assemblies. RADAR offers an excellent opportunity to improve the full Homestake LBNE project in physics reach, timeline, costs, and fostering international partnership. The anticipated duration of the project's construction is 5 years, with running happening between 2018 and 2023.}  

\newpage

\section{Introduction and Overview}

Neutrinos remain the least understood of the fundamental building blocks of matter and may very well provide answers to the most fundamental questions in our understanding of the evolution of the Universe. Tantalizingly, if violation of CP symmetry is observed in the neutrino sector, neutrinos may explain the matter-dominance observed in the present Universe, one of the deepest unsolved physics puzzles of our age. The most recent measurements of $\theta_{13}$ by the Daya Bay and RENO reactor experiments~\cite{ref:dayabay, ref:reno}, as well as by the long-baseline experiments MINOS~\cite{ref:minos} and T2K~\cite{ref:t2k}, making possible the determination of the neutrino mass hierarchy by the NuMI Off-axis $\nu_e$ Appearance (\nova)~\cite{ref:nova2} experiment during this decade. The Long-Baseline Neutrino Experiment (LBNE)~\cite{ref:LBNE}, projected for the next decade, will be able to unambiguously resolve the neutrino mass ordering and probe CP violation in the lepton sector, telling us if neutrinos play a crucial role in generating an asymmetry between matter and antimatter in the very early Universe.

 In this Letter of Intent (LOI), we propose the R\&D Argon Detector at Ash River (RADAR) project, aimed at enhancing the full Homestake LBNE project in four critical sectors: physics reach; timeline; cost; and the development of international partnerships. LBNE is a planned high-statistics neutrino oscillation experiment that will use a new neutrino beam line at Fermilab. In its full configuration, LBNE will operate a 34\,kton liquid argon time-projection chamber (LAr TPC) at the 4850' level of the Homestake Mine in Lead, South Dakota~\cite{ref:LBNE}.

The RADAR project described herein consists of a 6\,kton LAr TPC sited in the existing \nova~Far Detector building at Ash River, Minnesota.  The detector would record neutrino interactions from the 2\,GeV narrow band neutrino beam from the recently upgraded 700\,kW NuMI source at Fermilab.  With this arrangement, results from RADAR would allow for further optimization of LBNE to maximize its physics program reach, while significantly extending the physics capabilities of the NuMI beam-based \nova~program operating at Fermilab and Ash River. 

 Since LBNE anticipates building the 34\,kton detector out of comparably sized modules, detector R\&D costs listed in this LOI can be 100\% saved from the LBNE project.  Indeed, we aim to have this construction effort be one-and-the-same with the ongoing LBNE detector effort.  Furthermore, by constructing the first ever multi-kiloton LAr TPC in advance of readiness at the Homestake site, substantial cost savings and risk removal are possible for the full LBNE project, as the final LBNE designs can incorporate lessons learned from this single-module ``prototype'' at Ash River. The programmatic benefits are a key motivation for RADAR, given the high total project cost for the full LBNE project (\$1.44B). RADAR also allows the decoupling of some upfront R\&D and design costs from the full LBNE project, factorizing the funding load across separate projects. In addition, the readiness of the Ash River site allows detector construction and physics operations to begin sooner than at Homestake. The immediate site availability enables smaller initial cost commitments from international partners and will keep the long-baseline LAr program vibrant through this decade.

In the sections below we describe the physics potential of RADAR, discuss the programmatic synergies through which RADAR supports the full-scale LBNE effort, and give a detailed breakdown of the \$159M cost for the project including how much of this cost may represent direct savings for LBNE.

\section{Physics Reach}
The key motivations for RADAR are cost savings, schedule acceleration, and development of international collaborations for the full LBNE program, while allowing for optimization of the LBNE physics program and significant improvements to the physics reach of the existing \nova~program. To demonstrate the potential physics improvements we carried out sensitivity studies within the \nova code framework where we assumed:
\begin{itemize}
\item NuMI operates at 700 kW in the medium energy beam configuration.
\item The RADAR liquid argon detector has a fiducial mass of 5 kton and begins operation in 2019, running for 5 years.
\item \nova{} has a fiducial mass of 14 kton and operates for 10 years, including 6 years from 2014 to 2019 for its nominal run, and 4 years from 2020 to 2023, concurrently with RADAR data taking.
\item The T2K long-baseline experiment completes its projected 6-year run during this decade.
\end{itemize}
We further assume equal running time in neutrino and antineutrino modes for \nova{} and RADAR, but exclusive neutrino-mode running for T2K.

The NOvA contribution is calculated using the information in Ref.~\cite{ref:NOVACC}.  The T2K contribution is calculated using the information in Ref.~\cite{ref:T2KCC}.  The sensitivities from a LAr detector at Ash River are taken as equivalent to using NOvA detector technology with three times the mass of the LAr detector~\cite{novawhitepaper}.

\begin{figure}[!htbp]
\centering
\includegraphics[width=0.48\textwidth]{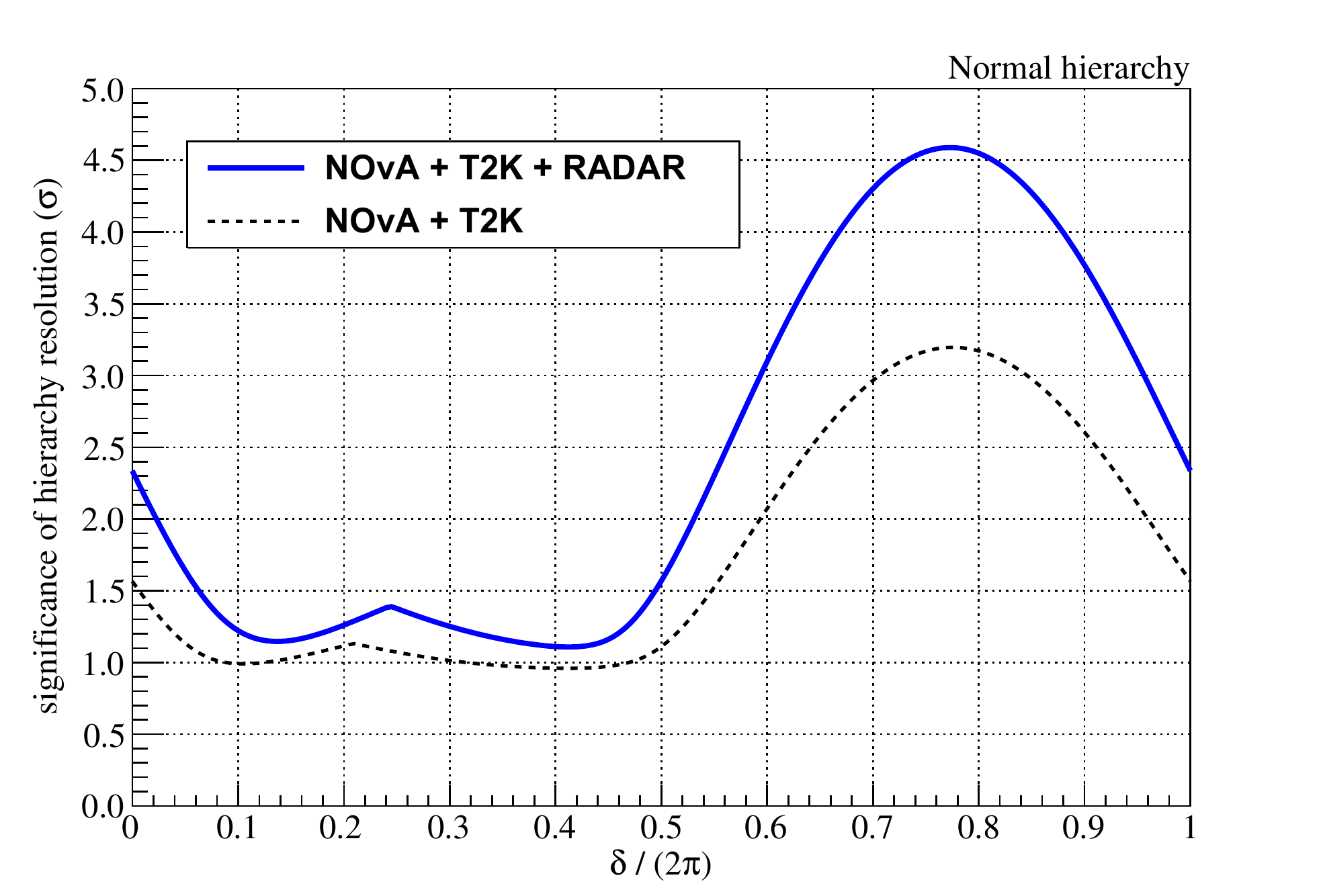}
\includegraphics[width=0.48\textwidth]{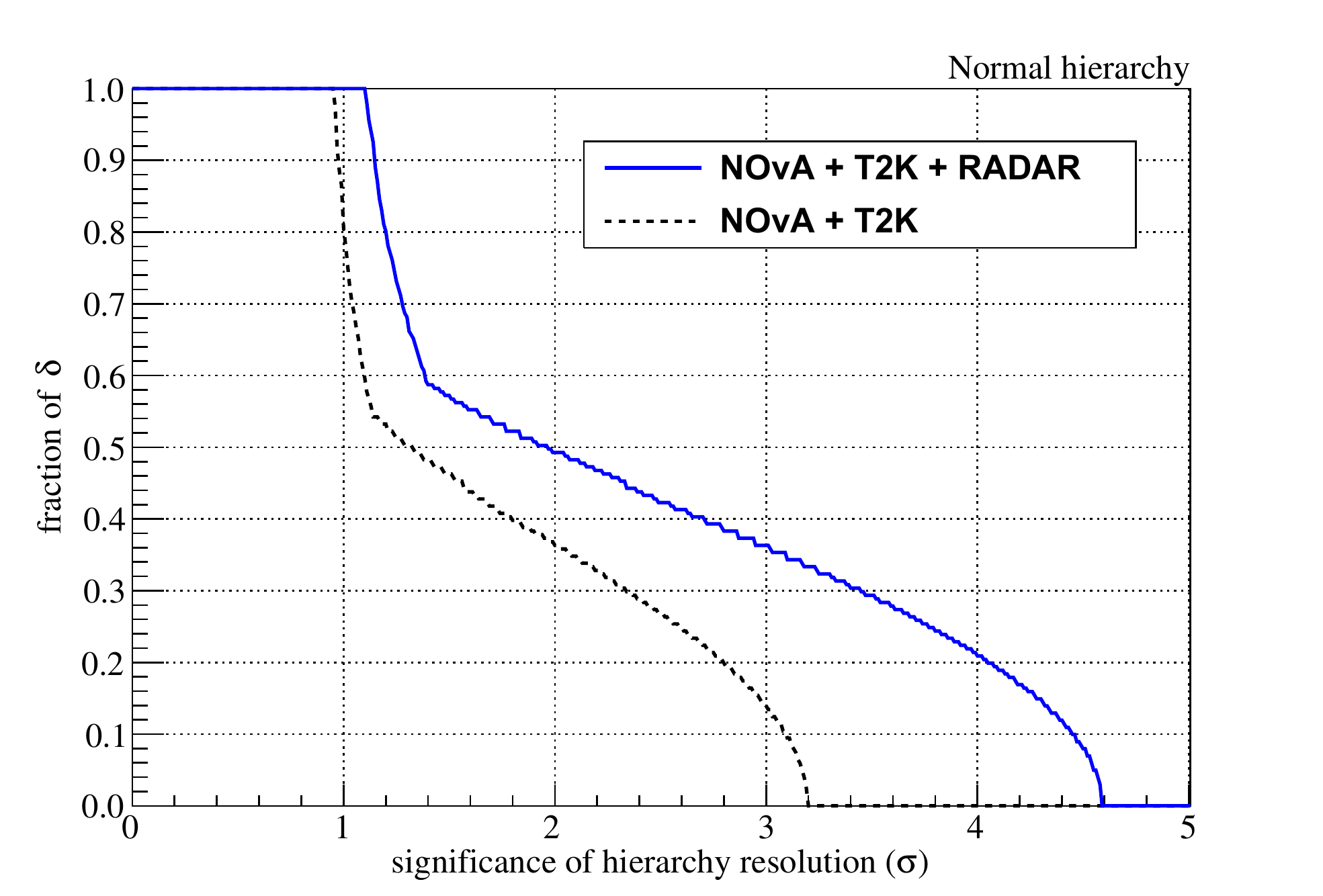}
\caption{\label{fig:hier} The significance of hierarchy resolution for \nova+T2K alone (black dashed) and with RADAR added (blue solid), shown both as a function $\delta$ (left) and in terms of the fraction of $\delta$ values covered at a given confidence level.  Normal hierarchy and maximal mixing are used.}
\end{figure}

\begin{figure}[!htbp]
\centering
\includegraphics[width=0.48\textwidth]{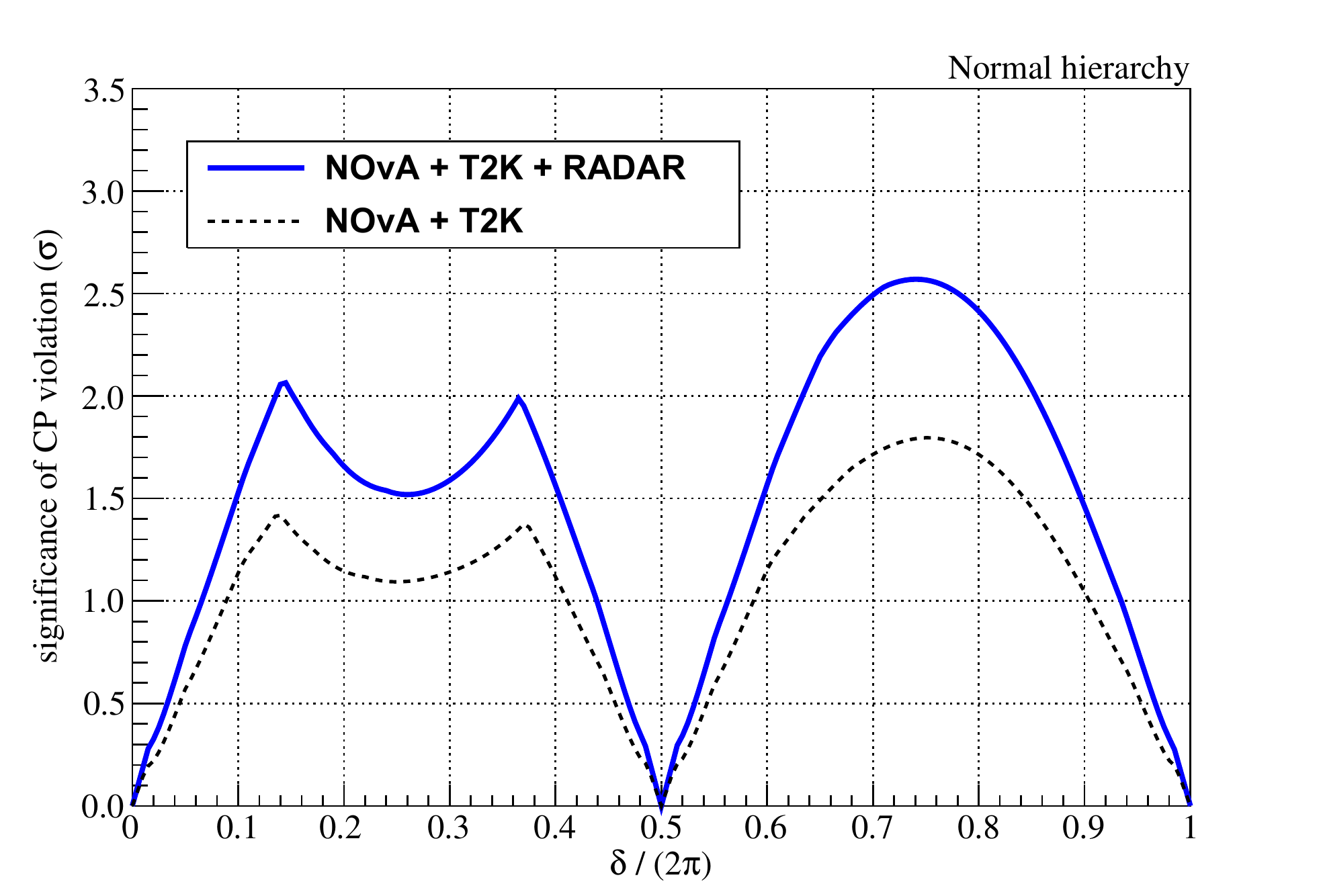}
\includegraphics[width=0.48\textwidth]{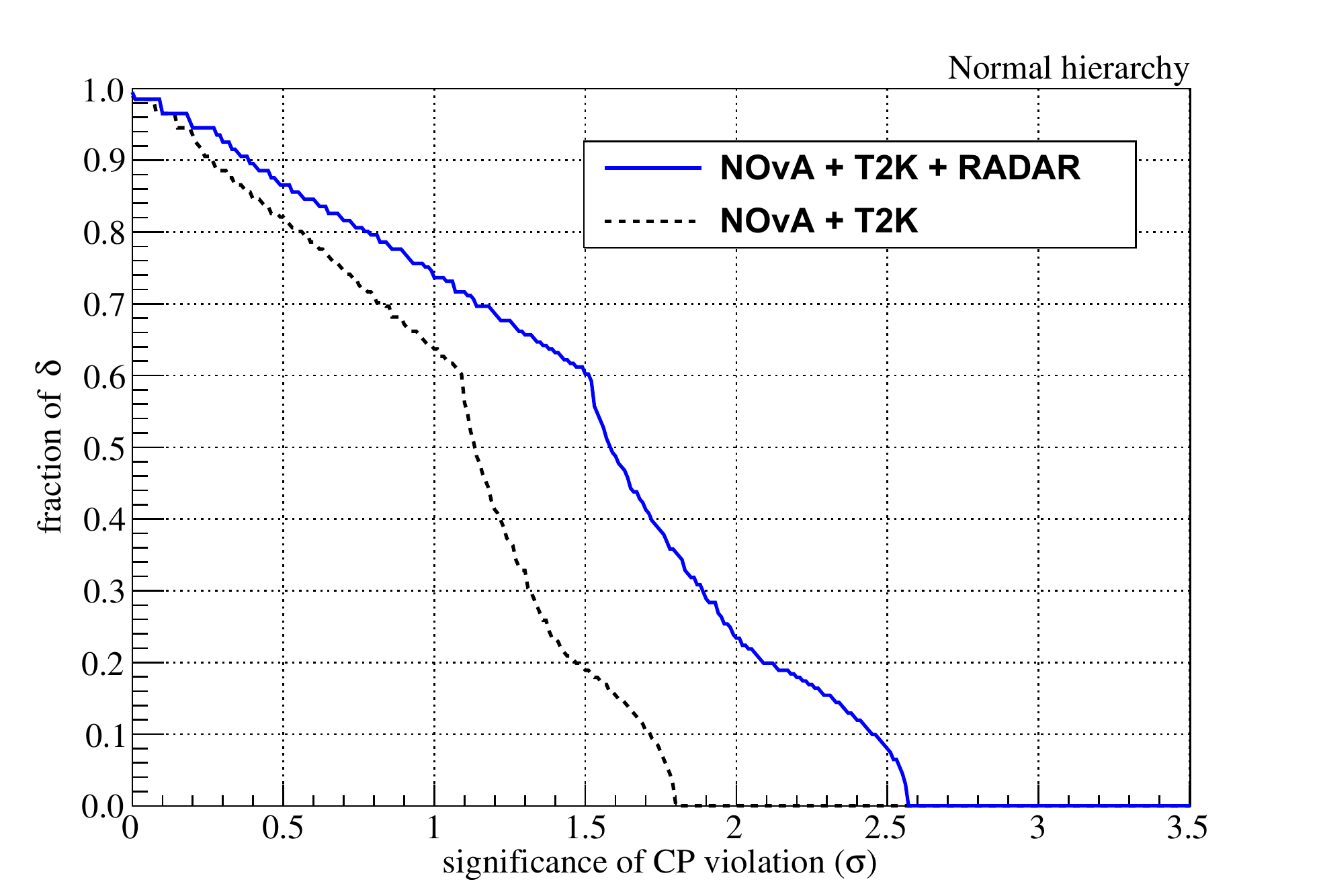}
\caption{\label{fig:cpv} The significance of CP violation for \nova+T2K alone (black dashed) and with RADAR added (blue solid), shown both as a function $\delta$ (left) and in terms of the fraction of $\delta$ values covered at a given confidence level.  Normal hierarchy and maximal mixing are used.}
\end{figure}

Figures~\ref{fig:hier} and \ref{fig:cpv} demonstrate the significant enhancement of the mass hierarchy and CP violation reach of the Ash River program provided by RADAR.  The hierarchy significance surpasses 4.5$\sigma$ at the most favorable $\delta$ values and is greater than 95\% C.L.\ for over half the $\delta$ range.  In the other half of the $\delta$ range, the hierarchy information gained from RADAR is correlated with the $\delta$ information. The reach for CP violation is also improved, with RADAR offering the first ever 95\% C.L.\ sensitivity to CP violation, in this case across a quarter of the $\delta$ range.

\begin{figure}[!tbp]
\centering
\includegraphics[width=0.6\textwidth]{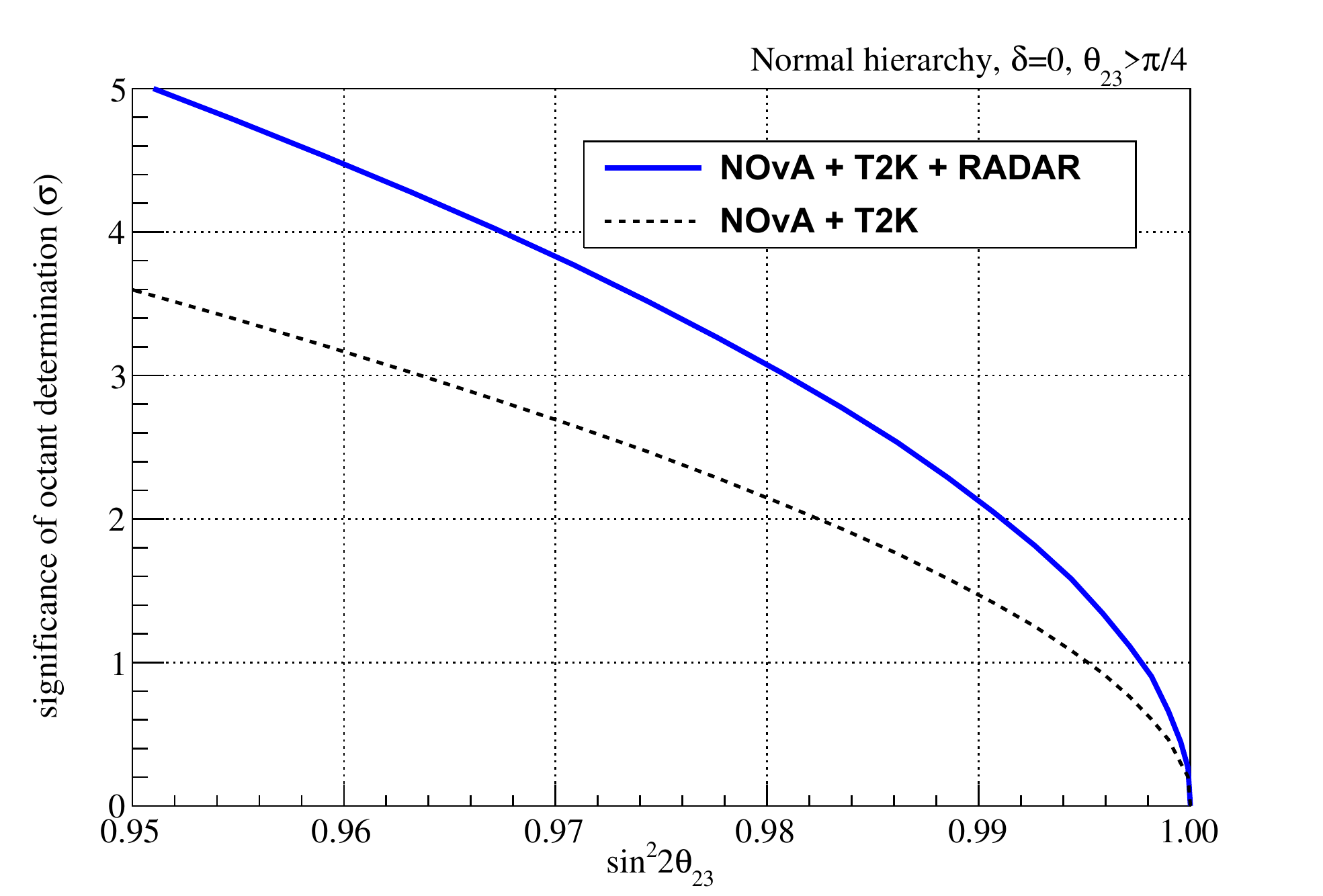}
\caption{\label{fig:octant} The significance with which the $\theta_{23}$ octant can be determined as a function of the true value of $\sin^22\theta_{23}$, shown for \nova~+~T2K alone (black dashed) and with RADAR added (blue solid).  $\theta_{13}$ is assumed known to 5\% and the uncertainty on $\sin^22\theta_{23}$ is taken from \nova~\cite{ref:NOVACC} and T2K projections~\cite{ref:T2KCC}. Normal hierarchy, $\delta=0$, and true $\theta_{23}>0$ are assumed. RADAR reduces the inaccesible range of $\sin^22\theta_{23}$ for a given desired confidence level by approximately a factor of two.}
\end{figure}

The near-maximal mixing implied by the current measurements of $\theta_{23}$ can also be tested at a level well beyond current limits.  RADAR will reduce the uncertainty on $\sin^22\theta_{23}$ by 40\% relative to \nova{} and can see deviations from maximal mixing at the 95\% C.L.\ all the way up to $\sin^22\theta_{23}=0.994$.  Furthermore, the flavor structure of the $\nu_3$ state, characterized by the octant of $\theta_{23}$, can be probed in RADAR beyond $\sin^22\theta_{23}=0.99$ at 95\% C.L., as shown in Fig.~\ref{fig:octant}.

\section{Integration with LBNE Program}
RADAR offers a platform to advance LBNE technology R\&D, develop international partnerships, and begin real physics analyses with a large LAr TPC.  We list here several key benefits to the full-scale LBNE program.
\begin{itemize}
\item The availability of the Ash River site allows construction of the first ever large (here, 6\,kton) LAr TPC module to happen in parallel with site preparation at Homestake.  As with any frontier-pushing detector technology, much will be learned when building this first detector. Collecting this valuable information prior to the start of detector construction at Homestake allows changes in design where beneficial for cost, installation ease, performance, etc.
\item To emphasize this point with an example:\ if construction of the Ash River module leads to design or construction procedure changes that provide 10\% cost savings per module, the resulting savings on the total 34\,kton detector cost, which is currently estimated at $\sim$ \$350M, would be significant. Combining these savings with the recycling of RADAR detector components, the total savings could be more than half of the RADAR project budget shown here.
\item The R\&D costs still pending for LBNE would be folded into RADAR, factorizing the costs across two separate projects. This would reduce the risk on the more cost-critical and larger LBNE project.
\item By collaborating with the RADAR program, international partners can research and take leadership of potential systems to be used in LBNE. They will also contribute to real physics deliverables sooner and at more palatable initial costs.
\item If the LBNE schedule were to slip relative to current collaboration projections, RADAR will keep technician and physicist expertise in the U.S.\ active. In such scenario, RADAR will generate LAr detector construction and operation experience, while producing forefront neutrino oscillation physics.
\end{itemize}

\section{Detector}
The proposed detector design will follow, wherever possible, the one resulting from R\&D carried out for LBNE~\cite{ref:LBNE}. Where necessary, for instance in optimizing use of the available space at the Ash River~building, the design will evolve within the RADAR program, but reusability of detector components and design will be a top priority.

\begin{figure}[thbp]
\centering
\includegraphics[width=0.6\textwidth]{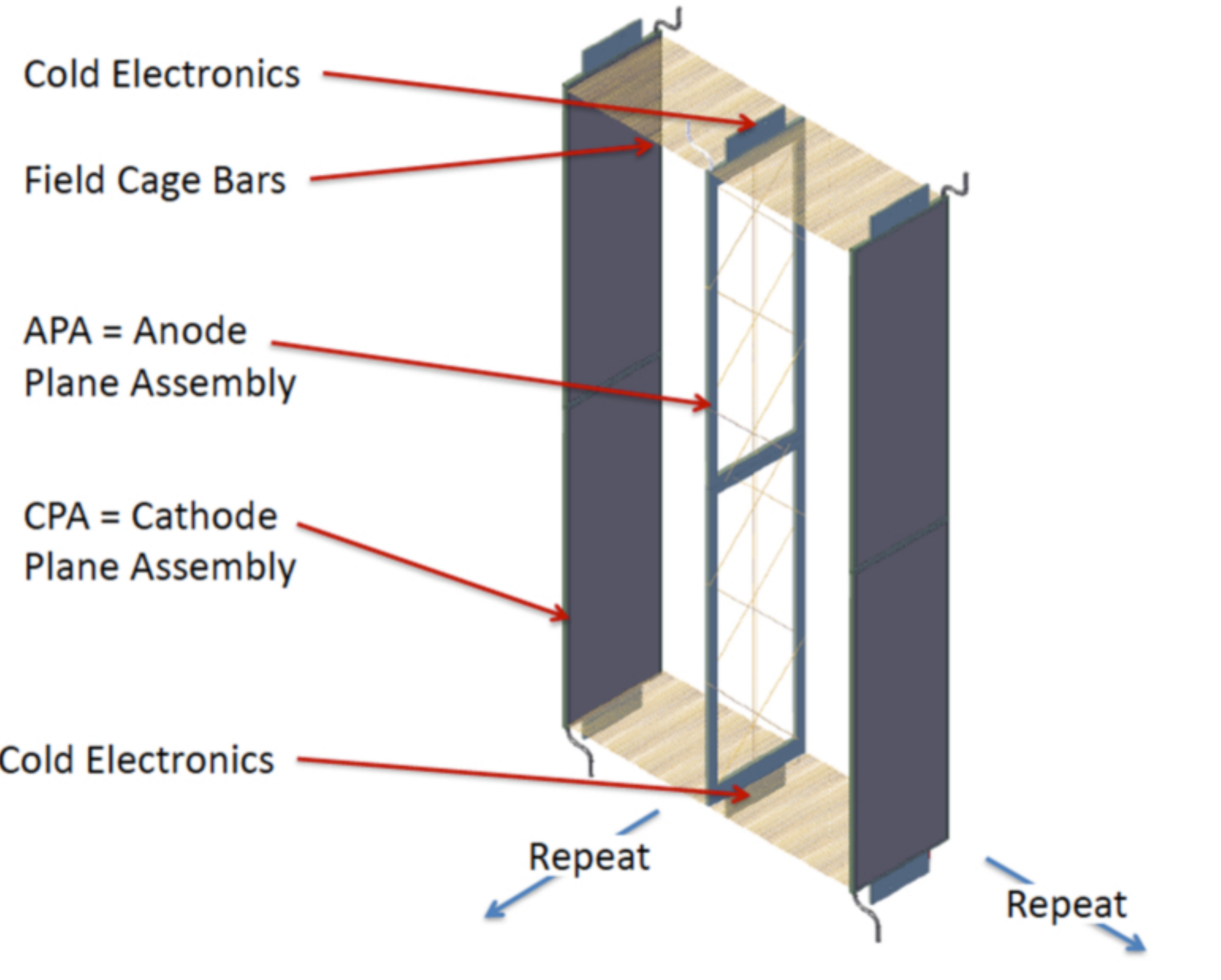}
\caption{\label{fig:assembly} Schematics of one assembly showing the APA, the two CPAs and the FCPs. The proposed LAr TPC design requires 84 of these assemblies, arranged in 14 longitudinal blocks of $3\times2$ (transverse$\times$vertical) assemblies.}
\end{figure}

The RADAR detector is a TPC composed of Anode Plane Assemblies (APAs), Cathode Plane Assemblies (CPAs), and Field Cage Panels (FCPs) to shape the uniform electric field of 500 V/cm between APAs and CPAs. The APAs carry the charge sense wires and scintillation light detection system and are instrumented with cold electronics. The CPAs provide the high voltage electrode that creates the uniform drift field. APAs, CPAs and FCPs are deployed inside a membrane cryostat containing ultra high-purity LAr maintained by a cryogenic system. The electrical and mechanical connections to the detector are routed via feedthroughs at the top of the cryostat. An illustration of one assembly of an APA, CPAS and FCPs is shown in Fig.~\ref{fig:assembly}.

\subsection{Location} 
\begin{figure}[thbp]
\centering
\includegraphics[width=0.9\textwidth]{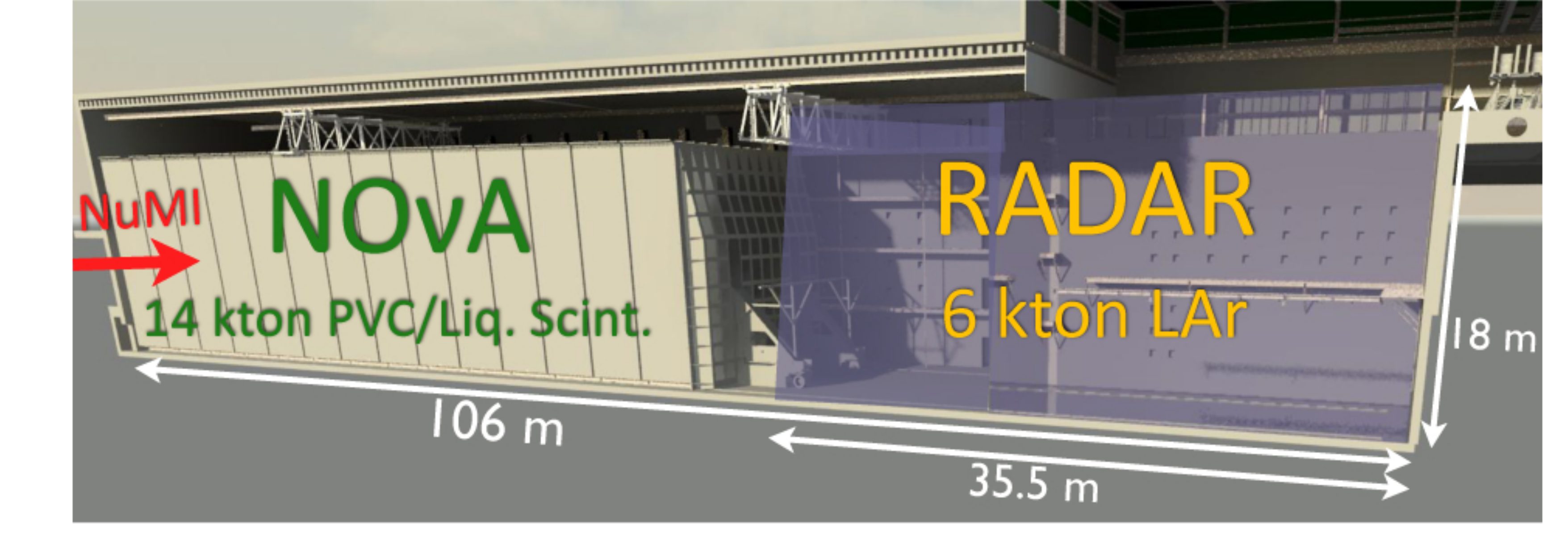}
\caption{\label{fig:location} Relative placement of the proposed RADAR TPC with respect to the \nova Far Detector at the Ash River building.}
\end{figure}

Under the assumption that \nova's Far Detector will remain at the CD-4 milestone of 14~kton mass, the available space for hosting the LAr volume inside the \nova building will include both the detector assembly area and the additional unoccupied space that was projected to harbor the originally proposed 18\,kton Far Detector. The dimensions of the total available space are approximately \unit[38]{m} x \unit[18]{m} x \unit[20]{m} (Length x Width x Height). A depiction of the RADAR TPC inside the Ash River building is shown in Fig.~\ref{fig:location}. The Ash River Building will require modifications to harbor the TPC, with additional power required and a bookend structure to hold the south side of the cryostat. As in the LBNE design, the remaining sides of the cryostat can be supported against the walls of the Ash River building, which are below ground level over the full height of the proposed cryostat. The cost of adding these enhancements to the building is estimated at \$10M.

\subsection{Detector Dimensions}
Following closely the design proposed for LBNE's LAr TPC, with a \unit[2.3]{m} drift distance and using APAs and CPAs with height increased from 7 to \unit[9]{m}, the proposed volume of the active detector is \unit[35.5]{m} x \unit[13.8]{m} x \unit[18]{m}. This volume takes into account structural support, insulation and secondary containment of the liquid argon. The resulting total LAr mass is \unit[6.0]{kton}. Assuming the same \unit{1.5}{m} longitudinal and \unit{0.3}{m} transverse fiducial cuts adopted in the LBNE design, the detector's fiducial mass is  \unit[4.6]{kton}. The total LAr TPC cost is estimated at \$159M, from scaling LBNE's 35\,kton TPC costs.

\subsection{Membrane Cryostat and Cryogenics}
 A membrane cryostat would be used, as currently suggested for LBNE. The cryostat consists of a rectangular tank with a 1.2-2~mm thick stainless-steel primary liner containing liquid nitrogen. Given the smooth walls of the \nova building, external support of the membrane cryostat walls should be relatively simple. Besides the stainless-steel membrane, the LBNE cryostat design requires a 30~cm layer of polyurethane insulation, a fiberglass-aluminum membrane for secondary LAr containment and a final external layer of 50~cm of polyurethane.
The cryogenic system will recycle the LAr volume and gradually increase its purity with each cycle. At a flow rate of 7~m$^3$/hr, similar to the ICARUS T600 TPC, a full recycling of the LAr volume is expected every 20 days. The LAr circulation will proceed through the top of the cryostat and the purifiers and re-condensers can be located in the receiving area of \nova's surface building. 
 We expect the cost of the cryostat and associated cryogenics to be approximately \$84.1M including LAr procurement and a 32\% scaled contingency based on LBNE estimates. The cost of the cryostat alone, including construction and contingency, is estimated at \$41M. The deployment of a storage tank to hold the argon between truck delivery and circulation into the TPC will cost an additional \$14.5M. 

\subsection{LAr TPC and Electronics}
  The active volume will be separated into $14\times3\times2=84$ drift regions, corresponding to 84 APAs. Each APA reads out 2560 channels, for a pitch of \unit[4.5]{mm} between wires.  The overall cost of the TPC is estimated at \$21M, including \$11M for production of APAs and CPAs, and \$5M for cold electronics. The electronics development can benefit from R\&D work already carried out for the MicroBooNE experiment~\cite{ref:microboone}, and also from ongoing R\&D work by LBNE. The estimated costs include a 49\% scaled contingency based on LBNE estimates. 

We note that the wire pitch of LBNE was designed with exquisite signal/background separation in mind, as the size of $\theta_{13}$ was unknown and the background mis-identification rate drove the $\sin^22\theta_{13}$ reach of LBNE. Given the now-known ``large'' value of $\theta_{13}$, the pressure on event identification is reduced somewhat, and the physics reach may be unaffected in any significant way with a reduced number of readout channels.  In developing the full proposal, we would explore this possibility for RADAR which could lead to significant cost reductions. In the present document, we assume the same \unit[4.5]{mm} wire pitch proposed for all LBNE LAr TPC options.

\subsection{LAr Photon Detector}
LBNE is planning to deploy a photon detection system to help mitigate cosmic ray backgrounds in accelerator physics analyses carried out with a detector on the surface. For underground detector deployment, this system can reduce proton decay backgrounds and improve detection of Supernovae bursts. 
Given the RADAR location at the surface and the key requirement of reusability of RADAR components in LBNE, a photon detection system is included in the cost budget. The proposed photon detection system uses 4 light guides layered into a paddle read out by a PMT. Each APA assembly would carry 10 of these paddles. With this configuration, RADAR provides a full-scale test stand for LBNE's photon detector design.

\subsection{Cost Summary and Reusability}
When the costs of modifying the Ash river building (\$10M),  installation and commissioning (\$11M), project management (\$14M), and of a LAr Photon detector (\$5M) are included, the total cost of the project is estimated at \$169M. The breakdown of these costs by project component is shown in Table~\ref{tab:costs}. The table also shows the cost of components that would be reused in LBNE in a favorable scenario: the LAr TPC assemblies and electronics would be fully reusable; the cryostat cannot be transferred from Ash River to Homestake, but the procured liquid argon (\$21M) and cryo-components such as pumps and condensers would be recyclable. In this scenario, we estimate \$21M of the cryostat and cryogenics costs may be offset by reusability, and a total \$58M, or about 34\% of the total budget, would count as direct savings for the LBNE project. At this juncture, we have not considered the potential difficulties of operating a LAr detector on the surface posed by large cosmic ray rates. Such studies are ongoing within the LBNE collaboration and will benefit greatly from input by the MicroBooNE experiment, scheduled to begin operations in 2014 at Fermilab. In the present design, close to 50\% of the RADAR TPC will be situated under the \nova overburden, a 3\,foot layer of barite and concrete (8.5\,mwe depth), while the remaining TPC volume will be at the assembly area with no shielding. If studies demonstrate the need to reduce rates of cosmic rate interactions, an extension of the overburden to the  assembly area is possible. However, it will require reinforcing the assembly area section of the roof of the \nova~building, adding civil construction costs estimated between \$20M and \$30M.

\begin{table}[!htbp]
\begin{tabular}{|l| S[table-format=3.2,table-number-alignment=center] |c| S[table-format=3.2,table-number-alignment=center] |}
 \hline
\text{Project Component} & \text{Scaled DIC Cost}  & \text{Included Scaled} & \text{Recoverable} \\ 
 & \text{\,\,\,\,\,\,\,\,\,\,($\times$\$1M)} & \text{Contingency} & \text{Costs ($\times$\$1M)}  \\ \hline
 \text{Ash River Bldg. Enhancements} & 10.0  & 50\% & 0.0 \\ \hline
\text{LAr TPC} & 25.6  & 49\% & 25.6 \\ \hline
\text{Cryogenics and Cryostat} & 84.1 & 32\% & 25.0 \\ \hline
\text{LAr Storage Tank} & 14.5 & 50\% & 0.0 \\ \hline
\text{Detector Monitoring and Control} & 4.6 & 50\% & 2.3 \\ \hline
\text{LAr Photon Detector} & 5.0 & 39\% & 5.0 \\ \hline
\text{Installation and Commissioning} & 11.2 & 52\% & 0.0 \\ \hline
\text{Project Management} & 14.0 & 16\% & 0.0 \\ \hline
\text{Total Cost} & 169.0 & 36\% & 57.8 \\ \hline
\end{tabular}
\caption{Anticipated scaled direct, indirect, and contingency costs for the RADAR project broken down by each project component. The costs and scaled contingency are estimated with the same cost spreadsheet used by LBNE to assess  the LAr detector option costs~\cite{ref:lbnecosts}. The estimated costs that could be recovered by reusing components in the final LBNE design are listed in the right-most column.}
\label{tab:costs}
\end{table}

\section{Project Timeline}
Given the experience with design and construction of MicroBooNE and LBNE, we anticipate inception to completion of the project would take 5 years. Building and operating RADAR within a schedule consistent with the projected timeline for the LBNE \unit{10}{kton} TPC may be challenging. However, including a 2 year schedule contingency, CD-4 approval for beginning of LBNE's data taking is scheduled for 2026 and the stretching of funding profiles in large projects is a perennial possibility. In such scenarios, RADAR provides a steady flow of scientific results during the upcoming decade and represents an excellent tool to train the next generation of neutrino physicists. Since we expect to build extensive synergies with the LBNE R\&D program, a detailed project timeline for RADAR will be developed in coordination with the LBNE project management.

\section{Summary}
By assembling the first multi-kiloton LAr TPC and placing it in the high-intensity NuMI beam line, RADAR can optimize the LBNE physics program reach and significantly enhance both the \nova physics program and ongoing LAr R\&D efforts. Valuable knowledge gained from LAr TPC construction and operations with neutrino interactions, along with the reusability of components in the LBNE full-scale Far Detector, can significantly offset the projected costs of the RADAR proposal. RADAR provides continuity of a vibrant US long-baseline neutrino physics program, potentially improving the sensitivity to the mass hierarchy to more than 4.5\,$\sigma$ at the most favorable $\delta$ and the reach for CP violation to 95\% C.L.\ across a quarter of the $\delta$ range. 

\onecolumn

\end{document}